\documentclass[11pt,twoside]{article}
\usepackage{CAGN2016}
\usepackage{graphicx}
\usepackage{hyperref}

\usepackage[T1]{fontenc} 

\usepackage{latexsym}
\usepackage{verbatim}

\begin{document}

\newcommand{\ion}[2]{#1~{\sc #2}}
\newcommand{\oii}{[\ion{O}{ii}]}
\newcommand{\oiii}{[\ion{O}{iii}]}
\newcommand{\nii}{[\ion{N}{ii}]}
\newcommand{\ha}{H$\alpha$}
\newcommand{\hb}{H$\beta$}

\vskip 1.0cm
\markboth{C. Morisset}{Photoionization models of CALIFA \ion{H}{ii} regions}
\pagestyle{myheadings}
%
%
\vspace*{0.5cm}
\parindent 0pt{Invited Review} 


\vspace*{0.5cm}
\title{Photoionization models of CALIFA HII regions. Genetic method.}

\author{Christophe Morisset}
\affil{Instituto de Astronomia \\ 
Universidad Nacional Autonoma de Mexico \\ 
email: {\tt chris.morisset@gmail.com} }

\begin{abstract}
I present recent and forthcoming works to model the CALIFA \ion{H}{ii} region using photoionization models. The first results are obtained with ad-hoc models (combining parameter determination by model fitting and strong line methods) while the next ones will use a Genetic Algorithm to fit the observations in a multi-dimensional space.
\end{abstract}

\section{Introduction}

The determination of the composition of \ion{H}{ii} regions is a challenge that has consequences on our understanding on the chemical evolution of our Universe, through the determination of the evolution of galaxies. Numerous methods have been developed during the last decades to determine chemical abundances. They are barely compatibles, especially when comparing strong line methods and photoionization model based methods (e.g. Fig. 2 of Kewley \& Ellison 2008).

In a recent paper, Morisset et al.(2016, hereafter Paper I) presented photoionization models for the CALIFA\footnote{Calar Alto Legacy Integral Field spectroscopy Area survey.} survey of \ion{H}{ii} regions (Sanchez et al. 2012) obtained with the Cloudy code (Ferland et al. 2013) driven by pyCloudy (Morisset 2013, 2014). The main idea is to obtain models for each one of $\sim$ 9,000 \ion{H}{ii} regions that reproduce the observed line ratio. We used the stellar population determined by the analysis of the underlying emission to compute the ionizing Spectral Energy Distribution (SED). The O/H abundance was determined from strong line method (from here the name of "hybrid models"). We finally kept only the ionization parameter U\footnote{The ionization parameter U being $Q_0 / 4.\pi.r^2.n_H.c$, where $Q_0$ is the number of ionizing photons emitted by the source per time unit, $r$ is the distance source-gas, $n_H$ is the hydrogen density and $c$ the light speed.} and the N/O abundance ratio as free parameters.

In the following I will first briefly describe the method used in Paper I and secondly concentrate on the genetic method that we will apply in a forthcoming work.

\section{Hybrid models from Paper I}
A photoionization code is a numerical way to transform a description of the object to be modeled (in terms of a set of input parameters including the SED shape and intensity, the gas distribution and composition and the description of the dust content) into a set of properties of the object that can be compared to observations: typically emission line intensities. Images, continuum spectra and absolute fluxes can also be considered.
This operation can be schematically represented by a link between a point from the multiple dimension parameter space (P-space) into the multiple dimensions observable space (O-space), see e.g. Morisset (2009). This transformation is highly no-linear, as seen for example when comparing the regular shape of a input Cartesian grid and the corresponding non regular grid in the O-space (See Fig. 2 from Stasinska et al. 2006, and Fig. 1 below). This implies that the reverse transformation (determination of the parameters values from the observations) is not trivial and that the solution may even not exist or not being unique.

When the number of observables is small, we need to reduce the number of free parameters (i.e. adapt the dimension of the P-space to the dimension of the O-space). This is for example achieved by hiding the stellar luminosity, the distance between the ionizing source and the ionized gas, and the gas density into the single variable U. 
In our modeling process of the CALIFA \ion{H}{ii} regions described in Paper I, we have only \nii, \oii, and \oiii\ emission line (with of course \ha\ and \hb, used to correct from reddening). We choose in a first step to only let log(U) and N/O as free parameters, and setting the O/H abundance using the diagnostic from Marino et al. (2013). The values of these two free parameters are obtained by fitting \nii/\ha, and \oiii/\hb\, while \oii/\hb\ is used as an {\it a posteriori} test to select the final set of successful models. Each region needs an ad-hoc process, as the ionizing SED is defined accordingly to the underlying corresponding stellar population. This leads to running close to 2 millions of models for only fitting two parameters; allowing a third parameter (O/H) to be free would increase this number to an indecent value for the computational power we currently have. At the end of the process, we obtain $\sim$ 3,200 ad-hoc models that fit the 3 line ratios simultaneously. All these models are store and publicly available from the 3MdB database\footnote{\url{https://sites.google.com/site/mexicanmillionmodels/}} (Morisset et al. 2015).

New relations are obtained between observables and parameters. We found that our models predict the [\ion{O}{iii}]4363/5007 line ratio to be close to (and a little bit smaller than) the observed values from Marino et al. (2013). This result contrasts with Dopita et al. (2013) who obtain models hotter than the observations. We suspect the ionizing SED softness to be responsible for this discrepancy between the two kinds of models. Further investigations are on the way.
One of the most puzzling results we obtained is from the comparison between the observed and theoretical \hb\ equivalent widths. The discrepancies between observations and models are associated to photon leaking and neighbor contamination, see Paper I for more details.

\section{Finding solutions in a 3D space: Genetic method}

In a second work (Morisse et al., in prep.), we explore the full 3D parameter space defined by log(U), O/H and N/O, by fitting the \nii/\ha, \oii/\hb, and \oiii/\hb\ line ratio simultaneously. To converge to the solution(s) we apply a genetic method (sort of Evolutionary Algorithm) based on the work by Canto et al. (2009) and already described by Morisset (2009). 
The genetic algorithm is based on successive grids of models (generations) following two basic rules of selection and evolution (crossover and mutation) to go from one generation to the next one: 
\begin{itemize}
\item The selection is obtained by defining a distance to a target in the O-space (in our case the observed line ratios) and criteria that a model needs to fit to be selected. It can be a minimum distance (defining a selection hyper-volume in the O-space) or a maximum number of objects to be selected after sorting them by increasing distance.
\item The evolution phase applies to the previously selected models to obtained the parameters for the next generation. We choose to only apply mutations (no crossover), following Canto et al. (2009). The values of the free parameters are modified by applying a Mutation Operator; in our case adding random values to the parameters. This Operator corresponds to a shift of the point corresponding to the model. The shift is defined by a vector of random size and direction in the P-space. The size of the vector defines the hyper-volume of the P-space that will be explored. This volume is supposed to decrease at each the generation to refine the quality of the fit.

\end{itemize}

We will only apply this new procedure to a selected amount of regions (the representative ones that cover most of the O-space and for which the data are of the best quality). 

\begin{figure}  
\begin{center}
\hspace{0.25cm}
\includegraphics[width=10cm]{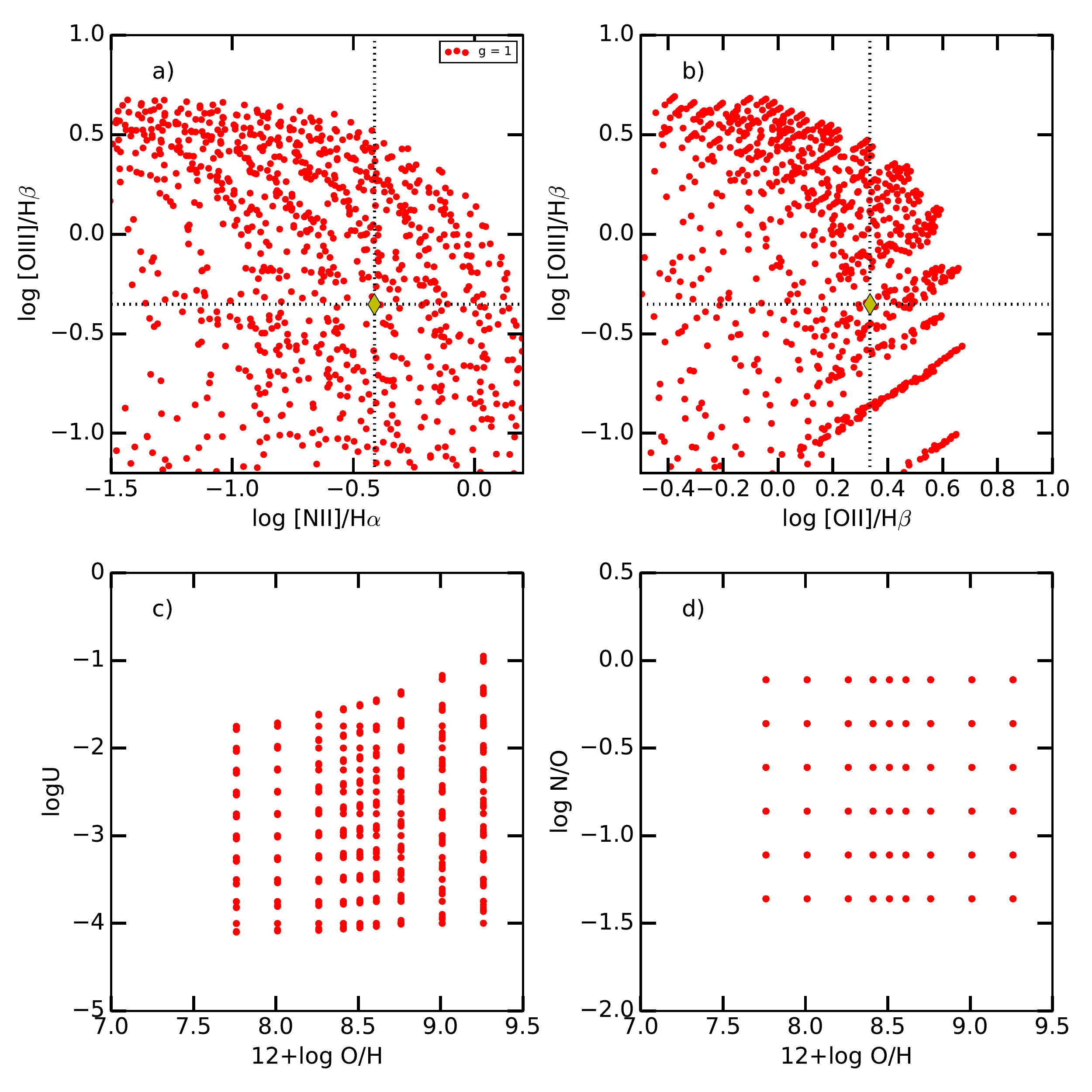}
\caption{Distribution of the models for the first generation. See text for details.}
\label{fig:g0}
\end{center}
\end{figure}

\begin{figure}  
\begin{center}
\hspace{0.25cm}
\includegraphics[width=10cm]{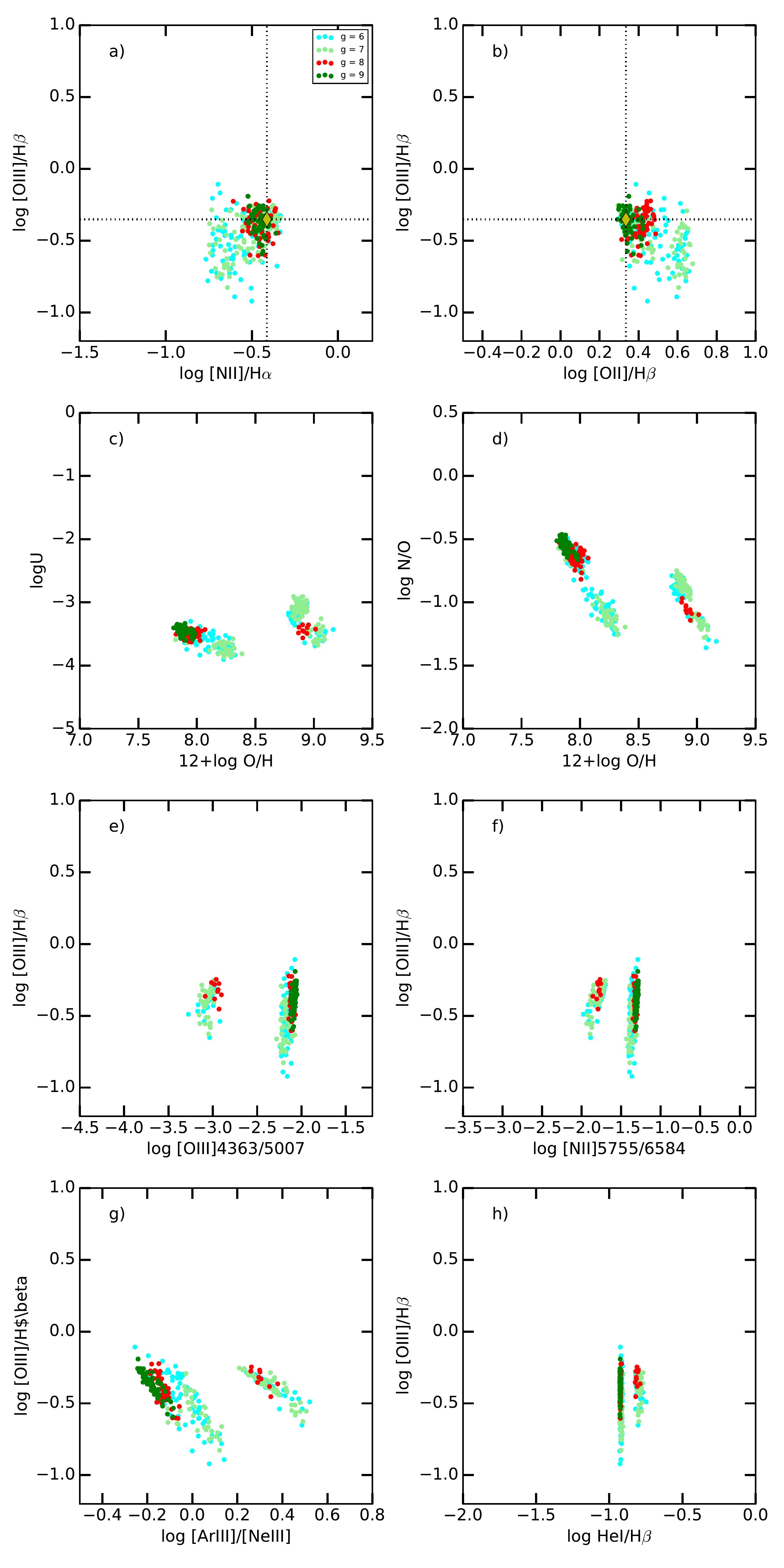}
\caption{Upper 4 panels: same as Fig.~\ref{fig:g0} for the generations 6, 7, 8 and 9. The lower 4 panels show predictions of the models in other dimensions of the O-space. See text for details.}
\label{fig:8}
\end{center}
\end{figure}

I present in Figs.~\ref{fig:g0} and \ref{fig:8} the process (in its current preliminary state) for the region 14 of NGC5947. 

In Fig.~\ref{fig:g0}, the red dots are the models of the first generation. The upper panels show the O-space projected on 2 planes: \oiii/\hb\ vs. \nii/\ha\ on the left and \oiii/\hb\ vs. \oii/\hb\ on the right; each red dot corresponds to the result of a model. The lower panels show the P-space projected on 2 planes: log(U) vs. O/H on the left and N/O vs. O/H on the right; each red dot corresponds to the values of the parameters for a model. The yellow diamond at the crossing of the two dotted lines on the upper panels corresponds to the observations of region 14 of NGC5947. We can appreciate the non linearity of the modeling process, considering the almost regular grid in the P-space, transformed into a messy grid in the O-space. Some {\it locii} in the O-space correspond to two points in the P-space due to the double value of the abundance for a given [\ion{O}{iii}]/\hb\ ratio.

In Fig.~\ref{fig:8}, the models of the generations 6 to 9 are presented, each generation corresponding to a different color. The models are converging to the position of the observed value symbolized by the yellow diamond at the crossing of the dotted lines (panels a and b). The panels c and d show the locations of the models in the P-space; the convergence is also clear. An interesting point here is that the results are bi-valuated: two clouds of different values in the P-space (especially for O/H) lead to the same locations in the O-space: this is the illustration of the degeneracy of the solution and the difficulties to determined O/H in lack of other information.
The 4 lower panels show the prediction of the models in other dimensions of the O-space, that could be compared to observations if available. Panels e and f show classical electron temperature diagnostics ([\ion{O}{iii}]$\lambda$4363/5007 and [\ion{N}{ii}]$\lambda$5755/6584 resp.). The two solutions differ by 1 and 0.5 dex respectively for these line ratios. But the auroral lines are almost impossible to observe for extra-galactic objects. The panel g shows the [\ion{Ar}{iii}]$\lambda$7135/[\ion{Ne}{iii}]$\lambda$3869 line ratio, used by the BOND method (Vale Asari et al. 2016) to differentiate the two branches of solutions. We see that this line ratio is discriminant by 0.5 dex; but we have to notice that this line ratio is very dependent on observational constraints (huge range of wavelength is necessary to obtain both lines at the same time). The \ion{He}{i}/\hb\ line ratio is mainly determined by the ionizing SED and is then the same for both solutions.

In our example, at generation 9 (green dots), one of the solutions is excluded from the process (the one corresponding to the higher O/H). It is actually an artifact of the method and will be corrected, as both solutions are valid at that stage. More astrophysical information is needed to choose the "right" solution, using for example N/O vs. O/H empirical relations.

\section{Conclusions}
Stay tuned for the paper II to appear.

\acknowledgments I would like to thank Oli Dors for this wonderful meeting. Many thanks to the editors for succeeding to join all our work in a coherent way. Thanks to Jorge Garcia-Rojas for reading the draft of this paper. And a very special thanks to Federico  for the {\it asado}.


\end{document}